\documentclass{jjap3}
\usepackage{txfonts}
\usepackage{multirow}
\usepackage{ulem}

\title{A frequency-stabilized light source at 399 nm using an Yb hollow-cathode lamp}
\author{Takehiko Tanabe$^{1}$\thanks{E-mail: t.tanabe@aist.go.jp}, Daisuke Akamatsu$^{1}$, Hajime Inaba$^{1}$, Sho Okubo$^{1}$, 
Takumi Kobayashi$^{1}$, Masami Yasuda$^{1}$, Kazumoto Hosaka$^{1}$, and Feng-Lei Hong$^{2,1}$}
\inst{$^{1}$National Metrology Institute of Japan (NMIJ), National Institute of Advanced Industrial Science and Technology (AIST), 
Tsukuba, Ibaraki 305-8563, Japan \\
$^{2}$Department of Physics, Graduate School of Engineering Science, Yokohama National University, Yokohama 240-8501, Japan}
\abst{
We demonstrate a diode laser system operating at 399 nm that is stabilized to the ${\rm 6s^{2}\ {^1}S_{0} - 6s6p\ {^1}P_{1}}$ 
electric dipole transition in ytterbium (Yb) atoms in a hollow-cathode lamp. 
The frequency stability of the laser reached $1.1 \times 10^{-11}$ 
at an averaging time of $\tau = 1\ \mathrm{s}$. 
We performed an absolute frequency measurement using an optical frequency comb and determined that 
the absolute frequency of the laser stabilized to the ${\rm {^1}S_{0} - {^1}P_{1}}$ transition 
in $^{174}\mathrm{Yb}$ was 751 526 522.26(9) MHz. 
We also investigated several systematic frequency shifts while changing some of the light source parameters 
and measured several isotope shifts. 
The measured laser frequency will provide useful information regarding the practical use of 
the frequency-stabilized light source at 399 nm. 
}

\begin{document}
\maketitle

\section{Introduction}

The frequency stabilization of lasers using atomic or molecular transitions is an essential technique 
for high precision laser spectroscopy and laser cooling experiments. 
Frequency-stabilized lasers are also important for many applications 
such as length and optical frequency metrology \cite{Udem2002,Hong2017}. 
There are, however, few frequency-stabilized lasers for use in the ultraviolet (UV) wavelength region. 
The development of a frequency-stabilized laser operating in the short wavelength region would provide 
several potential applications including in astrophysical spectroscopy 
thus making it possible to search exoplanets \cite{Wilken2012,Glenday2015}.

The wavelength and natural linewidth of the ${\rm 6s^{2}\ {^1}S_{0} - 6s6p\ {^1}P_{1}}$ electric dipole transition for ytterbium (Yb), 
which is a widely used atomic species in atomic physics experiments \cite{Takasu2003,Kohno2009,Noguchi2011,Nemitz2016,Nakajima2016}, 
are approximately 399 nm and 29 MHz, respectively.
The absolute frequency of this transition has been measured in several laboratories 
using an atomic beam method based on the fitting of the transition profile \cite{Das2005,Nizamani2010,Kleinert2016}, 
without locking the laser frequency to the center of the transition. 
To realize an optical reference in the UV region, it is useful to establish a frequency-stabilized laser 
using this transition in Yb. 
For example, a diode laser was stabilized to this transition using an Yb hollow-cathode lamp (HCL), 
which provides high number densities of atoms by using an electrical discharge \cite{Kohno2008,Wang2011}. 
In these experiments, Doppler-free isotope lines of Yb were observed using saturation absorption spectroscopy. 
Laser frequency stabilization using the observed Doppler-free lines was also performed. 
However, the absolute frequency of the stabilized laser has not been measured,  
and it may differ from the non-perturbed line center due to the interaction of Yb atoms with the buffer gas (Ne) in the HCL. 

In this study, we demonstrate a diode laser system operating at 399 nm that is stabilized to the 
electric dipole transition using Yb atoms in the HCL and measure the absolute frequency of the laser using an optical frequency comb. 
We observed improvements in the frequency stability by factors of 5 and 70, 
compared with the results obtained by Kohno $et\ al$. \cite{Kohno2008} and Wang $et\ al$. \cite{Wang2011}, respectively. 
Furthermore, we investigated laser frequency shifts when we changed some of the system parameters. 
The obtained laser frequency will be helpful information as regards the practical use of the 
frequency-stabilized light source at 399 nm.

\section{Experimental setup}
\label{sec:examples}

\subsection{Development of light source at 399 nm}

\begin{figure*}
\centering
\includegraphics[scale=0.57,bb=0 0 735 517]{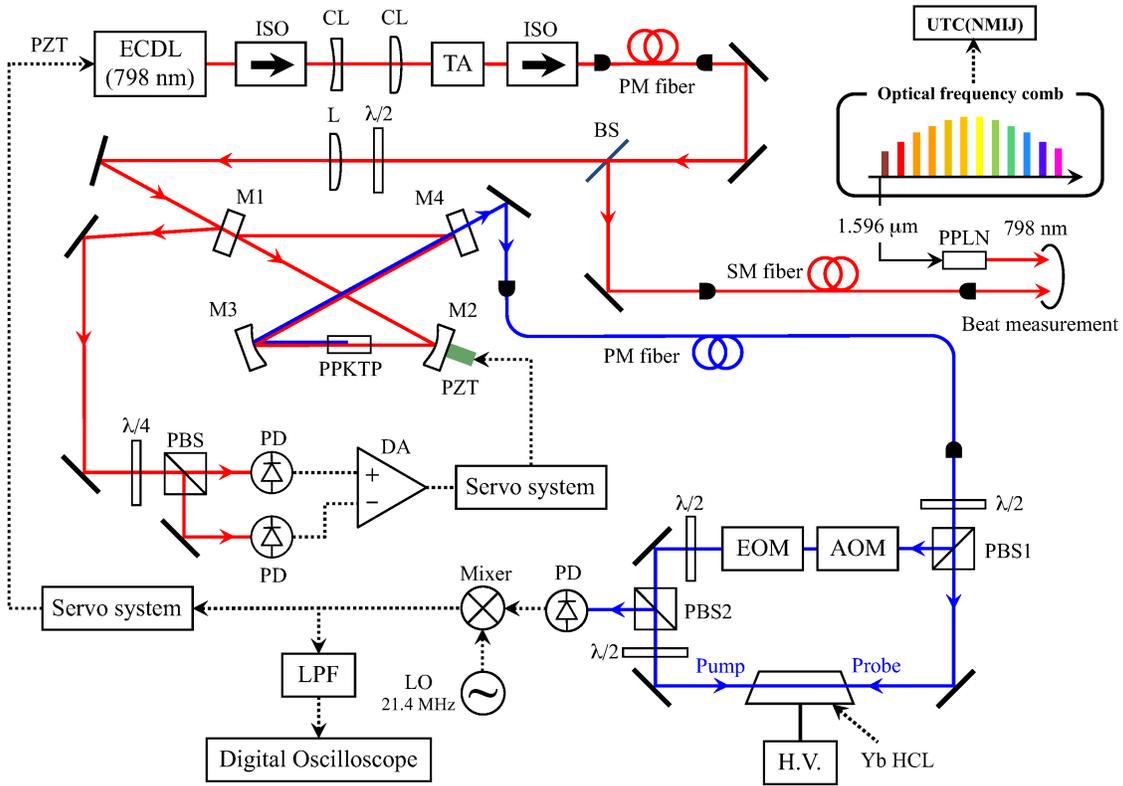}
\caption{(Color online) Schematic diagram of experimental setup. 
ECDL: external cavity diode laser, ISO: isolator, CL: cylindrical lens, TA: tapered amplifier, 
PM: polarization-maintaining, BS: beam splitter, $\lambda$/2: half-wave plate, L: lens, M: mirror, 
SM: single-mode, PZT: piezoelectric transducer, 
$\lambda$/4: quarter-wave plate, PBS: polarization beam splitter, 
PD: photodetector, DA: differential amplifier, 
AOM: acousto-optic modulator, EOM: electro-optic modulator, LO: local oscillator, 
PPKTP: periodically poled potassium titanyl phosphate, 
PPLN: periodically poled lithium niobate, HCL: hollow-cathode lamp, H.V.: high-voltage power supply, 
LPF: low-pass filter, UTC(NMIJ): coordinated universal time of the National Metrology Institute of Japan.
}
\label{fig:setup}
\end{figure*}

We have developed a light source at 399 nm based on the second harmonic generation (SHG) of a 798-nm diode laser. 
Figure \ref{fig:setup} shows a schematic diagram of the experimental setup.  
A home-built external cavity diode laser (ECDL) with a Littrow configuration operating at 798 nm was employed 
as a fundamental input light source.
A commercial diode laser chip (SANYO Electric, DL-LS2075) was used in the ECDL,  
and the length of the external cavity was about 10 mm.  
The typical output power of the ECDL at 798 nm was approximately 39 mW. 
Since this 798-nm power was too weak to generate sufficient power at 399 nm, 
we employed a home-built tapered amplifier (TA) using a laser diode chip (eagleyard Photonics, EYP-TPA-0795-02000-4006-CMT04-0000).  
The output beam of the ECDL passed through optical isolators with a total isolation of 60 dB 
and a pair of cylindrical lenses for spatial mode shaping and was then led into the TA. 
With a seed power of 39 mW, the TA provided a laser power about 400 mW at 798 nm. 
After a 30-dB isolator and several cylindrical lenses, 
the amplified 798-nm light was coupled into a polarization-maintaining (PM) fiber for spatial mode shaping. 
The output of the 798-nm light from the PM fiber was approximately 200 mW. 
Almost all the 798-nm light output from the PM fiber was sent to an enhancement cavity for frequency doubling. 
The remaining 798-nm light was sent to an optical frequency comb placed in another room for an  
absolute frequency measurement via a single-mode (SM) fiber. 

A 10-mm-long periodically poled potassium titanyl phosphate (PPKTP) crystal (Raicol Crystal) 
placed in an enhancement cavity with a bow-tie configuration was used for frequency doubling \cite{Han14}.
The PPKTP crystal was placed between two concave mirrors with a curvature of 50 mm to focus the fundamental light in the crystal. 
An antireflection coating for both fundamental and second harmonic lights was applied 
to both end facets of the PPKTP crystal. 
The enhancement cavity was approximately 378 mm long, which corresponds to a free spectral range of 793 MHz. 
The reflectivity of the input mirror M1 was 0.95 and those of the other mirrors M2, M3, and M4 exceeded 0.998 
for the fundamental light at 798 nm. 
The resonance frequency of the enhancement cavity was locked with the H{\"a}nsch-Couillaud scheme \cite{Hansch1980}. 
Error signals were generated with a differential amplifier (DA) and fed back 
to control the displacement of M2 through a piezoelectric transducer (PZT). 
Approximately 30 mW of blue light at 399 nm was obtained when 200 mW of fundamental light at 798 nm 
was coupled into the enhancement cavity. 
The temperature of the PPKTP crystal was kept at 49.4$^\circ$C. 
Light generated at 399 nm was coupled into a PM fiber for spatial mode shaping 
and approximately 10 mW of blue light at 399 nm was sent to a Yb spectrometer using the fiber.

\subsection{Frequency stabilization using an Yb hollow-cathode lamp}

Doppler-free spectroscopy of Yb was carried out based on the modulation transfer technique of saturation spectroscopy 
\cite{Shirley1982,Eickhoff1995,Hong2001,Hong2009}. 
One of the major advantages of this technique is that it generates dispersive-like absorption lineshapes, 
which sit on a nearly flat, zero background. 
A half-wave plate ($\lambda$/2 plate) and a polarization beam splitter (PBS, PBS1) were introduced 
to separate the pump and probe beams in the spectrometer and to adjust their power ratio. 
The pump beam was frequency shifted by 80 MHz using an acousto-optic modulator (AOM) 
and phase modulated by an electro-optic modulator (EOM) at a modulation frequency of 21.4 MHz. 
The AOM was introduced to prevent the interferometric baseline problems in the spectrometer 
that occur when using counter-propagating pump and probe beams. 
The phase modulation generated by the EOM was about 0.14 rad. 
The pump and probe beams were overlapped in the Yb hollow-cathode lamp (HCL, Hamamatsu Photonics), 
which was usually operated at a voltage of 171 V and a fixed current of 2 mA. 
The HCL driving voltage was optimized to maximize the signal-to-noise ratio of the modulation transfer signal, 
and the driving current was determined with a ballast resistor (22 k$\Omega$). 
The unmodulated probe beam developed sidebands inside the HCL as the result of a four-wave mixing process 
when saturation occurred \cite{Shirley1982}. 
The probe beam was separated from the pump beam by a PBS (PBS2) and detected by a photodetector (PD). 
A dispersive-like absorption signal was obtained by demodulating the signal from the detector 
using the signal from a local oscillator (LO). 
The waveform of the absorption signal was recorded with a digital oscilloscope 
after the signal had passed through a low pass filter (LPF). 
The cut-off frequency of the LPF used when recording the spectrum was 100 Hz.

\subsection{Frequency measurement system using an optical frequency comb}

In this study, the absolute frequency of the frequency-stabilized laser at 399 nm was obtained 
by measuring its fundamental light at 798 nm. 
The optical frequency comb used in the present measurements was home-built and based on a mode-locked 
erbium-doped fiber laser that covers a wavelength range of approximately 1 to 2 $\mu$m \cite{Inaba2006,Nakajima2010,Iwakuni2012}. 
The frequency comb was phase-locked to the coordinated universal time of the National Metrology Institute of Japan (UTC(NMIJ)). 
The comb component around 1596 nm was frequency doubled to 798 nm with a 20-mm-long single-pass 
periodically poled lithium niobate (PPLN) crystal (HC Photonics). 
The beat signals between the 798-nm fundamental laser light 
and the frequency-doubled comb components at 798 nm were detected and measured. 
The signal-to-noise ratio of the observed beat signals was approximately 30 dB at a resolution bandwidth of 300 kHz.

\section{Results}

\subsection{Frequency stability}
\begin{figure}
\centering
\includegraphics[scale=0.6,bb=0 0 605 417]{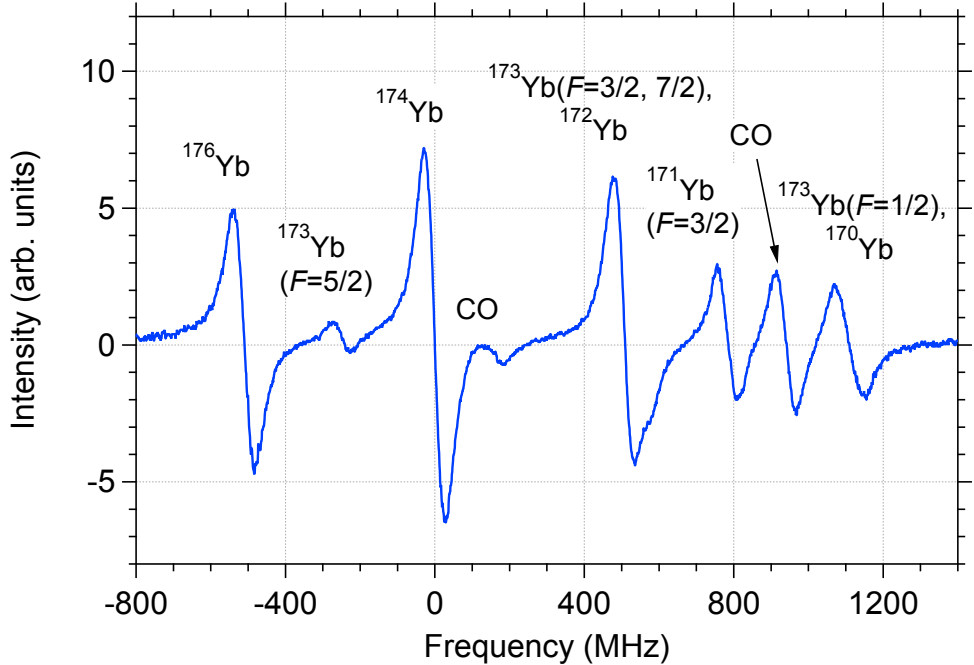}
\caption{(Color online) Modulation transfer signal of the ${\rm {^1}S_{0} - {^1}P_{1}}$ transition in Yb atoms in the HCL. 
The horizontal axis indicates the frequency difference between the scanning laser frequency 
and the observed $^{174}\mathrm{Yb}$ line center. 
The different isotopes and upper state hyperfine levels are labeled in the plot. 
$F$ and CO denote the total angular momentum quantum number and the cross-over resonance, respectively. 
}
\label{fig:Error_signal}
\end{figure}
Figure \ref{fig:Error_signal} shows the spectrum of the ${\rm {^1}S_{0} - {^1}P_{1}}$ transition 
in Yb atoms in the HCL observed with modulation transfer spectroscopy. 
The laser frequency was scanned by using the PZT of the ECDL. 
Every isotope except $^{168}\mathrm{Yb}$ was observed, 
because the natural isotopic abundance of $^{168}\mathrm{Yb}$ is relatively small (0.13 \%) compared with that of other isotopes.  
The spectrum widths shown in Fig. \ref{fig:Error_signal} are $60-70$ MHz, 
which are wider than the natural linewidth of the ${\rm {^1}S_{0} - {^1}P_{1}}$ transition in Yb atoms, 29 MHz. 
The broadening of the modulation transfer signal is considered to be caused by the pressure broadening 
that originates from the collision of Yb atoms with the buffer gas (Ne) sealed in the HCL. 
This has already been reported not only for Yb \cite{Kohno2008,Wang2011} but also for Ca \cite{Eble2007}.
Wang $et\ al$. \cite{Wang2011} reported that the width of the modulation transfer signal of the transition 
in Yb atoms in their HCL was about 80 MHz. 

\begin{figure}
\centering
\includegraphics[scale=0.7,bb=0 0 599 416]{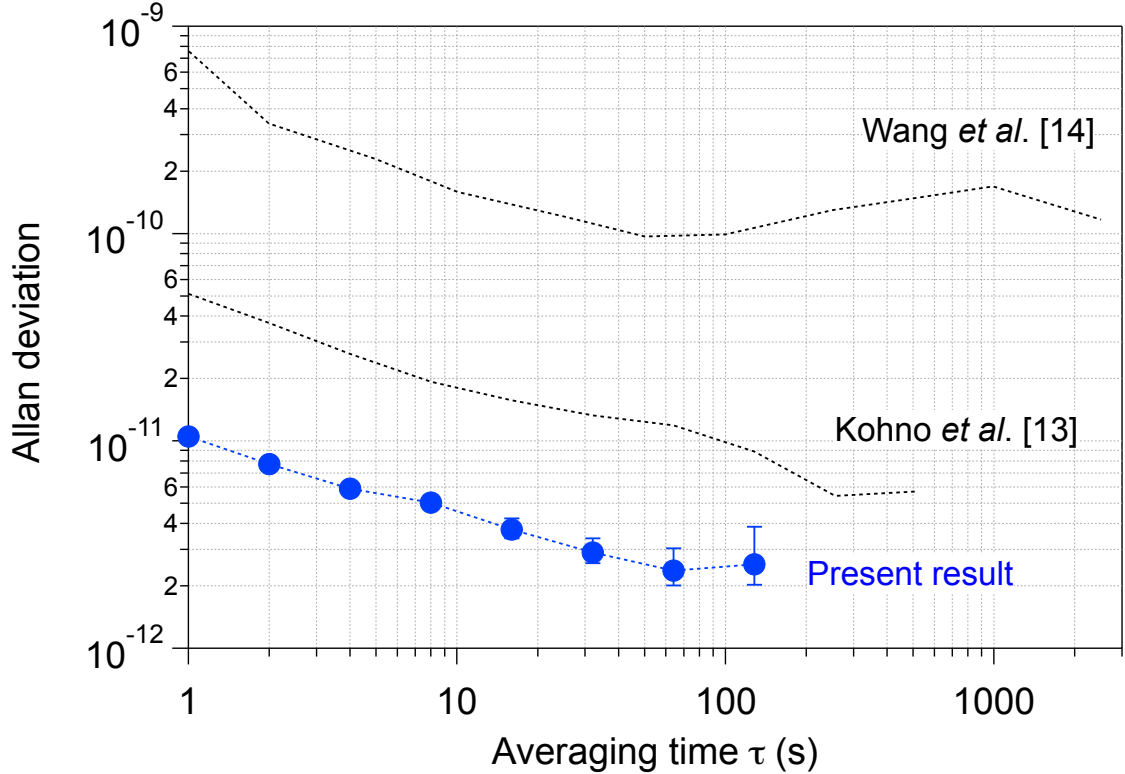}
\caption{(Color online) Typical Allan deviation calculated from the measured beat frequency between the ECDL 
stabilized to the ${\rm {^1}S_{0} - {^1}P_{1}}$ transition in $^{174}\mathrm{Yb}$ atoms in the HCL and the optical frequency comb.
For comparison, Fig. \ref{fig:Frequency_stability} also plots the results 
obtained by Kohno $et\ al$. \cite{Kohno2008} and Wang $et\ al$. \cite{Wang2011}.}
\label{fig:Frequency_stability}
\end{figure}

The laser frequency was stabilized to the center of the dispersive-like lineshapes 
by feeding back the absorption signal (error signal) to the PZT of the ECDL through a servo system. 
Figure \ref{fig:Frequency_stability} shows the Allan deviation calculated from the observed beat frequency 
between the laser stabilized to the $^{174}\mathrm{Yb}$ line in the HCL and the optical frequency comb. 
The obtained Allan deviation is $1.1 \times 10^{-11}$ at an averaging time of $\tau = 1\ \mathrm{s}$ 
and basically follows the $\tau^{-1/2}$ slope until $\tau = 60\ \mathrm{s}$. 
The Allan deviation of the UTC(NMIJ) is $10^{-13}$ at $\tau = 1\ \mathrm{s}$ \cite{Inaba2006}, 
which is much smaller than that of the present laser system. 
The frequency noise added by the optical frequency comb is $10^{-17} - 10^{-18}$ \cite{Millo2009}.  
Therefore, the observed stability of the beat frequency shown in Fig. \ref{fig:Frequency_stability} was limited 
by the frequency stability of the Yb-HCL-stabilized ECDL. 

For comparison, in Fig. \ref{fig:Frequency_stability} we also plot the laser frequency stability 
observed by Kohno $et\ al$. \cite{Kohno2008} and Wang $et\ al$. \cite{Wang2011}.  
We observed improvements in the frequency stability ($\tau = 1\ \mathrm{s}$) by factors of 5 and 70, 
compared with those in Ref. 13) and Ref. 14), respectively. 
We believe the improvement results from our use of a different spectroscopic method and frequency measurement system.

\subsection{Absolute frequency measurement}

\begin{figure}[ht]
\centering
\includegraphics[scale=0.6,bb=0 0 424 410]{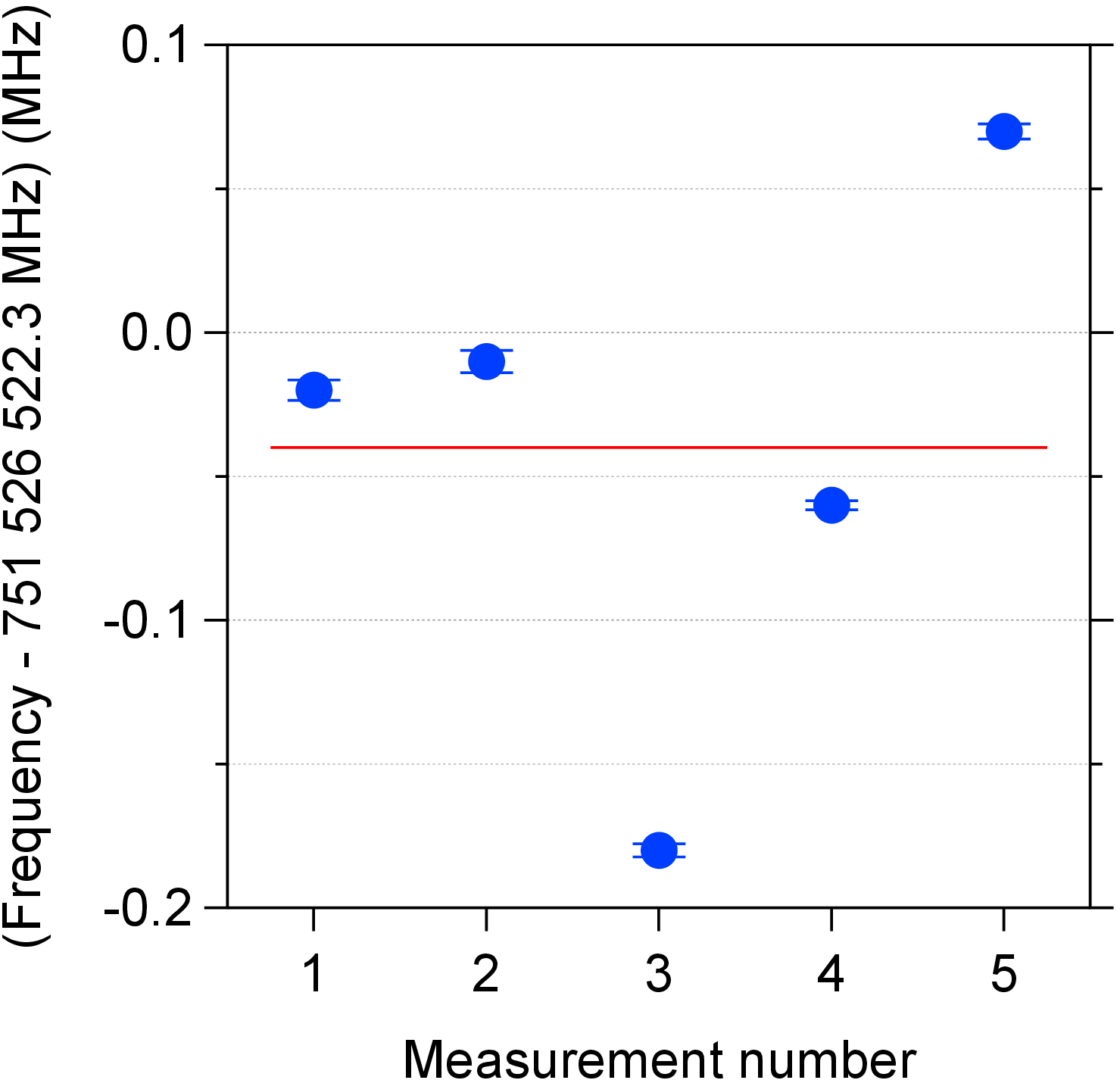}
\caption{(Color online) Absolute frequency measurement results for a laser operating at 399 nm 
stabilized to the $\rm{{^1}S_{0} - {^1}P_{1}}$ transition in $^{174}\mathrm{Yb}$ atoms in an HCL 
measured at the 2-mA driving current of the HCL, a pump power of 5 mW and a probe power of 1 mW. 
The solid red line indicates the average value of the five measured frequencies.}
\label{fig:Results_174Yb}
\end{figure}

Let $f_{\mathrm{beat}}$ be the beat frequency between the comb and the laser at 798 nm. 
The 798-nm light frequency $\nu$ is expressed as follows, 
\begin{eqnarray}
\nu = n f_{\mathrm{rep}} + f_{\mathrm{CEO}} + f_{\mathrm{beat}},  \label{eq:nu_798}
\end{eqnarray}
where $n$ is the mode number of the comb, and $f_{\mathrm{rep}}$ and $f_{\mathrm{CEO}}$ are the repetition rate 
and carrier-envelope offset frequencies, respectively. 
Both $f_{\mathrm{rep}}$ and $f_{\mathrm{CEO}}$ are phase-locked to UTC(NMIJ). 
The final quantity that must be determined is the mode number $n$ of the optical frequency comb. 
If the optical frequency to be measured has already been reported with an uncertainty much smaller than $f_{\mathrm{rep}}$,
$n$ can simply be determined by solving eq. (\ref{eq:nu_798}) for $n$.  
However, in the absolute frequency measurement of the ${\rm 6s^{2}\ {^1}S_{0} - 6s6p\ {^1}P_{1}}$ transition 
using the atomic beam method \cite{Das2005,Nizamani2010,Kleinert2016}, 
there is a considerable difference between the reported results (maximum of 660 MHz). 
Therefore, we determined the mode number of the comb by using two independent optical frequency combs 
with different $f_{\mathrm{rep}}$ values (88 and 123 MHz). 
This method is almost the same as that used in the study reported by Holzwarth $et\ al.$ \cite{Holzwarth2001}. 

Figure \ref{fig:Results_174Yb} shows the results of the absolute frequency measurement of the laser 
stabilized to the ${\rm {^1}S_{0} - {^1}P_{1}}$ transition in $^{174}\mathrm{Yb}$ atoms in the HCL. 
During the measurements, we repeated the process of locking and unlocking the laser. 
All the measurements were performed at an HCL driving voltage of 171 V, a pump power of 5 mW and a probe power of 1 mW.
Each measurement in Fig. \ref{fig:Results_174Yb} was calculated from 600 to 1000 beat frequency data, 
where each frequency datum was measured by using a frequency counter with a gate time of 1 s. 
The uncertainty bars in Fig. \ref{fig:Results_174Yb} were given by the Allan deviation at the longest averaging time 
among our measurements and are hard to see because the uncertainties are very small. 
The average value of the results shown in Fig. \ref{fig:Results_174Yb} was 
\begin{eqnarray*}
751\ 526\ 522.26(9)\ \mathrm{MHz}. \label{eq:nu_174}
\end{eqnarray*}
The standard deviation of the five measured frequencies was 0.09 MHz. 
We also investigated the systematic frequency shifts.

\subsection{Systematic shifts}

\begin{figure}[t]
\centering
\includegraphics[width=\columnwidth,bb=0 0 616 553]{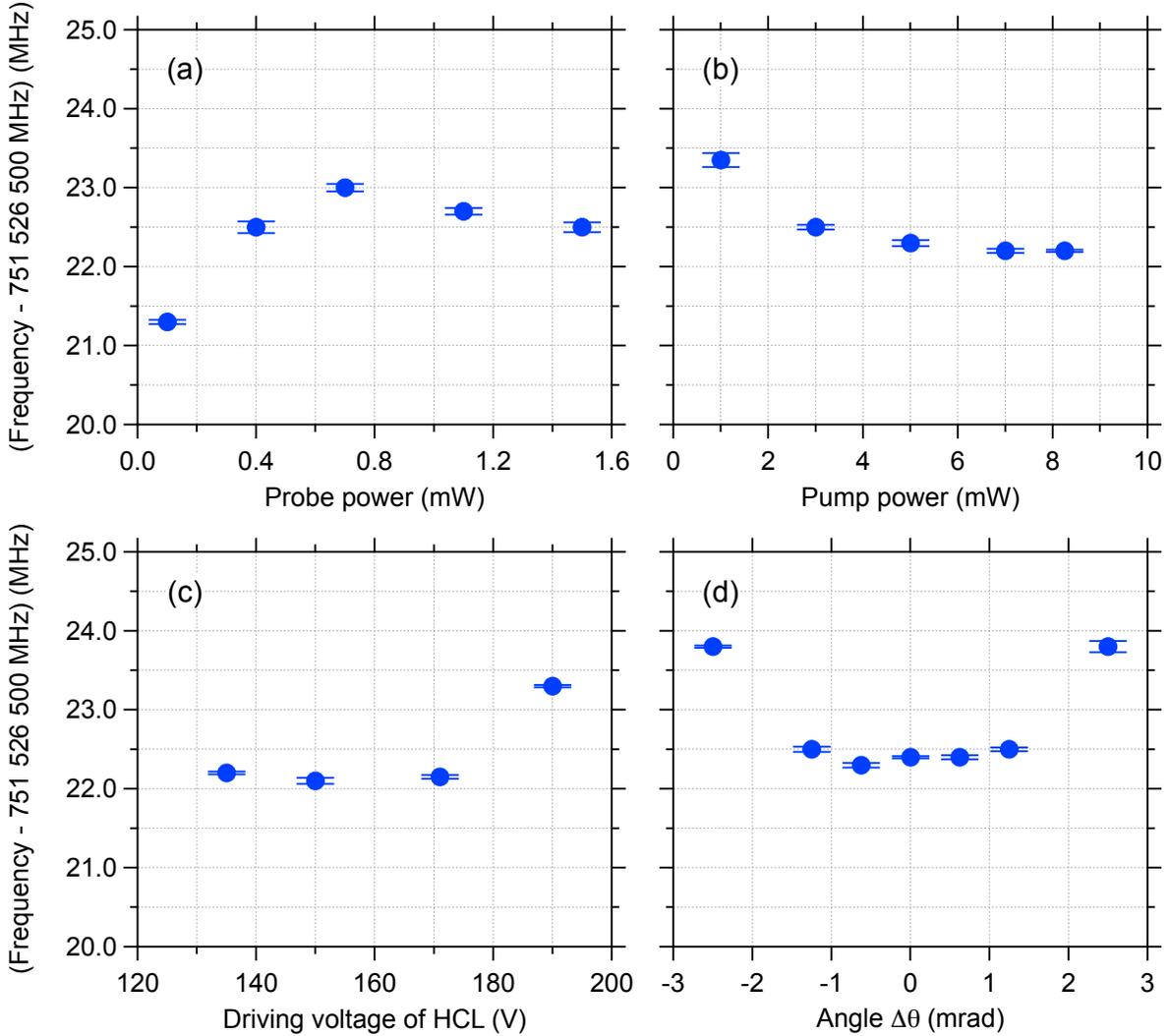}
\caption{(Color online) Frequency shifts of the Yb-HCL-stabilized ECDL: (a) probe power shift, (b) pump power shift, 
(c) driving voltage of the HCL and (d) misalignment of pump and probe beams. 
All measurements were carried out when the ECDL was stabilized to the ${\rm {^1}S_{0} - {^1}P_{1}}$ transition 
in $^{174}\mathrm{Yb}$ atoms in the HCL.}
\label{fig:systematic_shifts}
\end{figure}

Several systematic frequency shifts of the Yb-HCL-stabilized ECDL were investigated. 
All the measurements were carried out when the ECDL was stabilized to the ${\rm {^1}S_{0} - {^1}P_{1}}$ transition 
in $^{174}\mathrm{Yb}$ atoms in the HCL.
Figures \ref{fig:systematic_shifts} (a) and (b), respectively, show the frequency shift 
measured as a function of the probe power in the $0.1-1.5$ mW range 
and as a function of the pump power in the $1-8.3$ mW range. 
In Figs. \ref{fig:systematic_shifts} (a) and (b), maximum frequency shifts of up to $0.5$ MHz were observed 
except in the low pump and probe power regions. 
We also measured the frequency shift as a function of the HCL driving voltage. 
As shown in Fig. \ref{fig:systematic_shifts} (c), maximum frequency shifts of up to 0.2 MHz were observed 
in the $135-170$ V range and the estimated frequency shift was less than 6 kHz/V. 
Recently, Jang $et\ al.$ reported that their laser frequency strongly depended 
on the HCL driving voltage (approximately 15 MHz/mA) \cite{Jang2014}, 
whereas we observed no clear dependence in our measurements. 
The misalignment of the pump and probe beams induces a frequency shift \cite{Hong2004}. 
We checked this effect by varying the pointing direction of the pump beam,  
and the results are shown in Fig. \ref{fig:systematic_shifts} (d). 
Let $\varDelta \theta$ be the angle between the pump and probe beams. 
At $\varDelta \theta = 0$ mrad, the pump and probe beams overlapped in the HCL. 
A maximum frequency shift of 0.3 MHz was observed until $\varDelta \theta = \pm\ 2.5$ mrad. 
This indicates that the frequency shift caused by the possible misalignment of the pump and probe beams 
during the measurements was negligibly small.
In addition, the servo electronics offset, i.e., the dc voltage offset between the baseline of the spectrum and the lock point, 
was adjusted to below 3 mV to avoid an offset in the laser frequency. 
Since the amplitude of the error signal was typically 15 V (peak-to-peak), 
this dc offset level corresponds to a frequency shift of less than 6 kHz. 
Furthermore, we observed no obvious frequency shifts for phase modulation depths ranging from 0.072 to 0.14 rad.

Considering the above results, the observed statistical uncertainty in Fig. \ref{fig:Results_174Yb} cannot be fully explained 
by the investigated systematic effects. 
In other words, the statistical uncertainty is dominant compared with the systematic uncertainty 
under the operational coverage of our diode laser system. 
There may be systematic effects that are unknown or not taken into account in the present measurement, 
for example the residual amplitude modulation (RAM) in the phase modulation 
caused by the etalon effect of the EOM \cite{Whittaker1985}. 
According to its specifications, the EOM used in the present experiment works in a $500 - 900$ nm wavelength range. 
This means that there could be a rather large reflection of the 399-nm light 
at both ends of the EO crystal thus introducing a strong etalon effect. 
Furthermore, the etalon effect varies with the EO crystal temperature, 
and that could also be affected by long-term variations in the room temperature. 
We note here that the RAM may also be linked to the fact that the long-term frequency stability 
could not be improved when the average time exceeded $\tau = 60\ \mathrm{s}$. 

\subsection{Isotope shifts}

The absolute frequencies of the laser stabilized to other isotopes were also measured, 
and the results are summarized in Table \ref{tab:frequency_shifts}. 
Table \ref{tab:frequency_shifts} shows isotope frequency shifts relative to $^{174}\mathrm{Yb}$ in MHz units. 
The frequency uncertainties of the isotope frequency shifts (0.13 MHz) are calculated 
from the uncertainty of the absolute frequency measurement (0.09 MHz $\times \sqrt{2}$).
We were unable to stabilize the laser frequency to the line of the ${\rm ^{1}S_{0}}(F=1/2) - {\rm ^{1}P_{1}}(F=5/2)$ transition in 
$^{173}$Yb atoms, because the amplitude of the error signal was small as shown in Fig. \ref{fig:Error_signal}. 
The observed isotope frequency shifts of the ${\rm {^1}S_{0} - {^1}P_{1}}$ transition in $^{176}$Yb 
and the ${\rm ^{1}S_{0}}(F=1/2) - {\rm ^{1}P_{1}}(F=3/2)$ transition in $^{171}$Yb are in good agreement with previously reported values 
\cite{Deilamian1993,Loftus2001,Banerjee2003,Das2005,Wang2010,Nizamani2010,Johanning2011,Kleinert2016} 
within the uncertainties.
As for those lines involving multiple isotope transitions, our results are in good agreement with the results 
reported by Nizamani $et\ al.$ \cite{Nizamani2010}, 
but could not be compared with the results that resolve the isotope transitions by fitting the spectrum theoretically 
\cite{Deilamian1993,Loftus2001,Banerjee2003,Das2005,Wang2010,Johanning2011,Kleinert2016}. 
It is worth noting here that although the absolute frequencies of each isotope transition may be different 
in experiments using a HCL and an atomic beam, 
isotope frequency shifts are basically independent of the experimental methods.

\begin{table*}[t]
\centering
\caption{Frequency shifts of laser stabilized to certain points relative to $^{174}\mathrm{Yb}$ in MHz units. 
$F$ denotes the total angular momentum quantum number.}
\label{tab:frequency_shifts}
\begin{tabular}{cc}
\hline
Isotope and transition & Isotope frequency shift (MHz) \\
\hline
$^{176}$Yb, ${\rm {^{1}S_{0}} - {^{1}P_{1}}}$ & -509.63(13) \\ 
\hline
$^{173}$Yb, ${\rm ^{1}S_{0}}(F=1/2) - {\rm ^{1}P_{1}}(F=5/2)$ & Not stabilized \\ 
\hline
$^{174}$Yb, ${\rm {^{1}S_{0}} - {^{1}P_{1}}}$ & 0 \\
\hline 
$^{172}$Yb, ${\rm {^{1}S_{0}} - {^{1}P_{1}}}$ & \multicolumn{1}{c}{\multirow{3}{*}{534.80(13)}} \\ 
$^{173}$Yb, ${\rm ^{1}S_{0}}(F=1/2) - {\rm ^{1}P_{1}}(F=3/2)$       & \multicolumn{1}{c}{} \\ 
$^{173}$Yb, ${\rm ^{1}S_{0}}(F=1/2) - {\rm ^{1}P_{1}}(F=7/2)$       & \multicolumn{1}{c}{} \\ 
\hline
$^{171}$Yb, ${\rm ^{1}S_{0}}(F=1/2) - {\rm ^{1}P_{1}}(F=3/2)$   &  837.72(13) \\
\hline
$^{170}$Yb, ${\rm {^{1}S_{0}} - {^{1}P_{1}}}$   & \multicolumn{1}{c}{\multirow{2}{*}{1169.60(13)}} \\ 
$^{171}$Yb, ${\rm ^{1}S_{0}}(F=1/2) - {\rm ^{1}P_{1}}(F=1/2)$	     &\multicolumn{1}{c}{} \\ 
\hline
\end{tabular}
\end{table*}

\section{Discussion and conclusion}

\begin{figure}
\centering
\includegraphics[scale=0.5,bb=0 0 637 468]{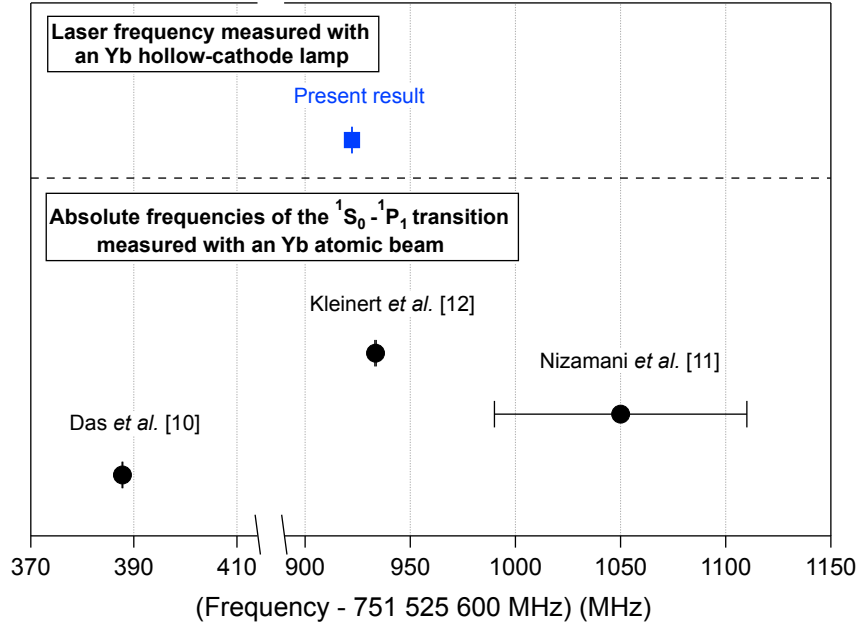}
\caption{(Color online) Comparison of the present result and the absolute frequencies 
of the ${\rm {^1}S_{0} - {^1}P_{1}}$ transition in $^{174}\mathrm{Yb}$ atoms 
measured at different laboratories \cite{Das2005,Nizamani2010,Kleinert2016}. 
The solid blue square shows the present result and the others were measured with an Yb atomic beam.
}
\label{fig:399nm}
\end{figure}

Figure \ref{fig:399nm} compares the measured laser frequency in the present experiment 
with the measured transition frequencies in the previous experiments. 
Das $et\ al.$ \cite{Das2005} and Nizamani $et\ al.$ \cite{Nizamani2010} measured 
the absolute frequency of the ${\rm {^1}S_{0} - {^1}P_{1}}$ transition in $^{174}\mathrm{Yb}$ atoms 
in an Yb atomic beam and found them to be 751 525 987.761(60) MHz and 751 526 650(60) MHz, respectively. 
More recently, Kleinert $et\ al.$ also reported the absolute frequency of the same transition in $^{174}\mathrm{Yb}$ atoms 
in an Yb atomic beam to be 751 526 533.49(33) MHz using an optical frequency comb \cite{Kleinert2016}. 
These three results do not agree with each other. 
Since Kleinert $et\ al.$ used an optical frequency comb in their frequency measurement \cite{Kleinert2016}, 
it is reasonable to consider that their result is more reliable than the earlier two measurements \cite{Das2005,Nizamani2010}. 
Our result for the laser frequency stabilized to the ${\rm {^1}S_{0} - {^1}P_{1}}$ transition in $^{174}\mathrm{Yb}$ atoms 
in the HCL is -11.23 MHz from Kleinert's value \cite{Kleinert2016}. 
In the HCL, Yb atoms interact with the buffer gas (Ne), 
while the Yb atomic beam is free from collisions with other atoms. 
Therefore, the difference between our result and that obtained by Kleinert $et\ al.$ \cite{Kleinert2016} 
can be mainly attributed to the pressure shift induced by collisions with the Ne buffer gas in the HCL.

The pressure broadening and shift of the ${\rm {^1}S_{0} - {^3}P_{1}}$ transition in Yb atoms at 556 nm 
have been extensively explored by Kimball $et\ al.$ \cite{Kimball1999}, 
while those for the ${\rm {^1}S_{0} - {^1}P_{1}}$ transition have yet to be reported.
We were informed by Hamamatsu Photonics that $5-10$ Torr Ne buffer gas has been sealed in the Yb HCL. 
Taking the center value of the buffer gas pressure (7.5 Torr), 
we calculated the pressure shift due to the Ne gas in the Yb HCL to be 1.5 MHz/Torr 
using the difference between our result and that obtained by Kleinert $et\ al.$ \cite{Kleinert2016}. 
Dammalapati $et\ al.$ \cite{Dammalapati2009} reported that the pressure shift due to the Ne buffer gas in the Ca HCL is 1.4 MHz/Torr. 
As for Sr atoms, Shimada $et\ al.$ \cite{Shimada2013} states that $5-10$ Torr of Ne buffer gas is sealed 
in the commercially available Sr HCL (Hamamatsu Photonics). 
Using the collisional shift values obtained by Chan and Gelbwachs \cite{Chan1992}, 
we calculated the pressure shifts caused by the Ne buffer gas in the Sr HCL to be 1.2 MHz/Torr. 
The pressure shifts of the Yb, Ca and Sr atoms caused by the Ne buffer gas are of the same order.

In conclusion, we stabilized a diode laser system, operating at 399 nm, 
to the ${\rm 6s^{2}\ {^1}S_{0} - 6s6p\ {^1}P_{1}}$ electric dipole transition in Yb atoms 
in an HCL and performed absolute frequency measurements using an optical frequency comb. 
We measured the absolute frequency of the laser stabilized to the ${\rm {^1}S_{0} - {^1}P_{1}}$ transition 
in $^{174}\mathrm{Yb}$ to be 751 526 522.26(9) MHz and also obtained the isotope frequency shifts 
relative to the $^{174}\mathrm{Yb}$ line center. 
We evaluated the systematic frequency shifts for different pump and probe powers, 
HCL driving voltages and alignments. 
The present results will provide a useful frequency reference at 399 nm.  
The developed frequency-stabilized laser at 399 nm can be used for laser cooling Yb atoms 
for atomic physics experiments such as those involving optical lattice clocks \cite{Kohno2009,Nemitz2016}, 
Bose-Einstein condensates \cite{Takasu2003} and quantum simulation \cite{Nakajima2016}.  
It is also useful to employ the frequency-stabilized laser 
as a frequency marker in an astro-comb \cite{Wilken2012,Glenday2015} covering 399 nm. 

\acknowledgments

We thank Dr. Tomonari Suzuyama of the National Metrology Institute of Japan (NMIJ), 
National Institute of Advanced Industrial Science and Technology (AIST) for his work in maintaining the UTC(NMIJ).
We dedicate this paper to the memory of Dr. Atsushi Onae of the NMIJ, AIST, who passed away while completing this manuscript. 
This work was supported by a Grant-in-Aid for Science Research 
(Grant Numbers JP25800231, JP24740281, JP23540472 and JP15K05238) from Japan Society for the Promotion of Science (JSPS). 


\begin{thebibliography}{10}

\bibitem{Udem2002}
T.~Udem, R.~Holzwarth, and T.~W. H{\"a}nsch: Nature {\bfseries 416} (2002) 233.

\bibitem{Hong2017}
F.-L. Hong: Meas. Sci. Technol. {\bfseries 28} (2017) 012002.

\bibitem{Wilken2012}
T.~Wilken, G.~L. Curto, R.~A. Probst, T.~Steinmetz, A.~Manescau, L.~Pasquini,
  J.~I. Gonzalez~Hernandez, R.~Rebolo, T.~W. H{\"a}nsch, T.~Udem, and
  R.~Holzwarth: Nature {\bfseries 485} (2012) 611.

\bibitem{Glenday2015}
A.~G. Glenday, C.-H. Li, N.~Langellier, G.~Chang, L.-J. Chen, G.~Furesz, A.~A.
  Zibrov, F.~K\"{a}rtner, D.~F. Phillips, D.~Sasselov, A.~Szentgyorgyi, and
  R.~L. Walsworth: Optica {\bfseries 2} (2015) 250.

\bibitem{Takasu2003}
Y.~Takasu, K.~Maki, K.~Komori, T.~Takano, K.~Honda, M.~Kumakura, T.~Yabuzaki,
  and Y.~Takahashi: Phys. Rev. Lett. {\bfseries 91} (2003) 040404.

\bibitem{Kohno2009}
T.~Kohno, M.~Yasuda, K.~Hosaka, H.~Inaba, Y.~Nakajima, and F.-L. Hong: Appl.
  Phys. Express {\bfseries 2} (2009) 072501.

\bibitem{Noguchi2011}
A.~Noguchi, Y.~Eto, M.~Ueda, and M.~Kozuma: Phys. Rev. A {\bfseries 84} (2011)
  030301.

\bibitem{Nemitz2016}
N.~Nemitz, T.~Ohkubo, M.~Takamoto, I.~Ushijima, M.~Das, N.~Ohmae, and
  H.~Katori: Nat. Photonics {\bfseries 10} (2016) 258.

\bibitem{Nakajima2016}
S.~Nakajima, S.~Taie, T.~Ichinose, H.~Ozawa, L.~Wang, M.~Troyer, and
  Y.~Takahashi: Nat. Phys. {\bfseries 12} (2016) 296.

\bibitem{Das2005}
D.~Das, S.~Barthwal, A.~Banerjee, and V.~Natarajan: Phys. Rev. A {\bfseries 72}
  (2005) 032506.

\bibitem{Nizamani2010}
A.~H. Nizamani, J.~J. McLoughlin, and W.~K. Hensinger: Phys. Rev. A {\bfseries
  82} (2010) 043408.

\bibitem{Kleinert2016}
M.~Kleinert, M.~E. Gold~Dahl, and S.~Bergeson: Phys. Rev. A {\bfseries 94}
  (2016) 052511.

\bibitem{Kohno2008}
T.~Kohno, M.~Yasuda, H.~Inaba, and F.-L. Hong: Jpn. J. Appl. Phys. {\bfseries
  47} (2008) 8856.

\bibitem{Wang2011}
W.-L. Wang, Y.~Jie, J.~Hai-Ling, B.~Zhi-Yi, M.~Long-Sheng, and X.-Y. Xu: Chin.
  Phys. B {\bfseries 20} (2011) 013201.

\bibitem{Han14}
Y.~Han, X.~Wen, J.~Bai, B.~Yang, Y.~Wang, J.~He, and J.~Wang: J. Opt. Soc. Am.
  B {\bfseries 31} (2014) 1942.

\bibitem{Hansch1980}
T.~W. H{\"a}nsch and B.~Couillaud: Opt. Commun. {\bfseries 35} (1980) 441.

\bibitem{Shirley1982}
J.~H. Shirley: Opt. Lett. {\bfseries 7} (1982) 537.

\bibitem{Eickhoff1995}
M.~L. Eickhoff and J.~L. Hall: IEEE Trans. Instrum. Meas. {\bfseries 44} (1995)
  155.

\bibitem{Hong2001}
F.-L. Hong, J.~Ishikawa, Z.-Y. Bi, J.~Zhang, K.~Seta, A.~Onae, J.~Yoda, and
  H.~Matsumoto: IEEE Trans. Instrum. Meas. {\bfseries 50} (2001) 486.

\bibitem{Hong2009}
F.-L. Hong, H.~Inaba, K.~Hosaka, M.~Yasuda, and A.~Onae: Opt. Express
  {\bfseries 17} (2009) 1652.

\bibitem{Inaba2006}
H.~Inaba, Y.~Daimon, F.-L. Hong, A.~Onae, K.~Minoshima, T.~R. Schibli,
  H.~Matsumoto, M.~Hirano, T.~Okuno, M.~Onishi, and M.~Nakazawa: Opt. Express
  {\bfseries 14} (2006) 5223.

\bibitem{Nakajima2010}
Y.~Nakajima, H.~Inaba, K.~Hosaka, K.~Minoshima, A.~Onae, M.~Yasuda, T.~Kohno,
  S.~Kawato, T.~Kobayashi, T.~Katsuyama, and F.-L. Hong: Opt. Express
  {\bfseries 18} (2010) 1667.

\bibitem{Iwakuni2012}
K.~Iwakuni, H.~Inaba, Y.~Nakajima, T.~Kobayashi, K.~Hosaka, A.~Onae, and F.-L.
  Hong: Opt. Express {\bfseries 20} (2012) 13769.

\bibitem{Eble2007}
J.~F. Eble and F.~Schmidt-Kaler: Appl. Phys. B {\bfseries 88} (2007) 563.

\bibitem{Millo2009}
J.~Millo, R.~Boudot, M.~Lours, P.~Y. Bourgeois, A.~N. Luiten, Y.~L. Coq,
  Y.~Kersal\'{e}, and G.~Santarelli: Opt. Lett. {\bfseries 34} (2009) 3707.

\bibitem{Holzwarth2001}
R.~Holzwarth, A.~Y. Nevsky, M.~Zimmermann, T.~Udem, T.~W. H{\"a}nsch, J.~von
  Zanthier, H.~Walther, J.~C. Knight, W.~J. Wadsworth, P.~S.~J. Russell, M.~N.
  Skvortsov, and S.~N. Bagayev: Appl. Phys. B {\bfseries 73} (2001) 269.

\bibitem{Jang2014}
G.~H. Jang, M.~Na, B.~Moon, and T.~H. Yoon: Phys. Rev. A {\bfseries 89} (2014)
  062510.

\bibitem{Hong2004}
F.-L. Hong, J.~Ishikawa, Y.~Zhang, R.~Guo, A.~Onae, and H.~Matsumoto: Opt.
  Commun. {\bfseries 235} (2004) 377.

\bibitem{Whittaker1985}
E.~A. Whittaker, M.~Gehrtz, and G.~C. Bjorklund: J. Opt. Soc. Am. B {\bfseries
  2} (1985) 1320.

\bibitem{Deilamian1993}
K.~Deilamian, J.~D. Gillaspy, and D.~E. Kelleher: J. Opt. Soc. Am. B {\bfseries
  10} (1993) 789.

\bibitem{Loftus2001}
T.~Loftus, J.~R. Bochinski, and T.~Mossberg: Phys. Rev. A {\bfseries 63} (2001)
  023402.

\bibitem{Banerjee2003}
A.~Banerjee, U.~D. Rapol, D.~Das, A.~Krishna, and V.~Natarajan: Europhys. Lett.
  {\bfseries 63} (2003) 340.

\bibitem{Wang2010}
W.~Wen-Li and X.~Xin-Ye: Chin. Phys. B {\bfseries 19} (2010) 123202.

\bibitem{Johanning2011}
M.~Johanning, A.~Braun, D.~Eiteneuer, C.~Paape, C.~Balzer, W.~Neuhauser, and
  C.~Wunderlich: Appl. Phys. B {\bfseries 103} (2011) 327.

\bibitem{Kimball1999}
D.~F. Kimball, D.~Clyde, D.~Budker, D.~DeMille, S.~J. Freedman, S.~Rochester,
  J.~E. Stalnaker, and M.~Zolotorev: Phys. Rev. A {\bfseries 60} (1999) 1103.

\bibitem{Dammalapati2009}
U.~Dammalapati, I.~Norris, and E.~Riis: J. Phys. B {\bfseries 42} (2009)
  165001.

\bibitem{Shimada2013}
Y.~Shimada, Y.~Chida, N.~Ohtsubo, T.~Aoki, M.~Takeuchi, T.~Kuga, and Y.~Torii:
  Rev. Sci. Instrum. {\bfseries 84} (2013) 063101.

\bibitem{Chan1992}
Y.~C. Chan and J.~A. Gelbwachs: J. Phys. B {\bfseries 25} (1992) 3601.

\end{thebibliography}

\end{document}